\documentclass[a4paper,11pt]{article}
\usepackage{pos}
\usepackage{mathtools} 
\usepackage{xcolor}
\usepackage{graphicx}
\usepackage{hyperref}
\usepackage{microtype}
\usepackage{cleveref}
\usepackage{booktabs}
\usepackage{xcolor}
\usepackage{tabu}
\usepackage{siunitx}

\title{Three-pion effects in $K^0-\bar{K}^0$ mixing}

\author*[a,b]{Andrew W. Jackura}
\author[a,b,c,d]{Ra\'ul A. Brice\~no}
\author[e]{Maxwell T. Hansen}

\affiliation[a]{Thomas Jefferson National Accelerator Facility,
	12000 Jefferson Avenue,
	Newport News, VA 23606, USA
}

\affiliation[b]{Department of Physics,
	Old Dominion University,
	Norfolk, Virginia 23529, USA
}
\affiliation[c]{Department of Physics, University of California, Berkeley, CA 94720, USA}    
\affiliation[d]{Nuclear Science Division, Lawrence Berkeley National Laboratory, Berkeley, CA 94720, USA}  
\affiliation[e]{Higgs Centre for Theoretical Physics, School of Physics and Astronomy,\\
The University of Edinburgh, Edinburgh EH9 3FD, UK}

\emailAdd{ajackura@jlab.org}

\abstract{The rate of mixing between a neutral kaon and an anti-kaon ($K^0-\bar{K}^0$) is given, in part, by a long-range matrix element, defined with two insertions of the weak Hamiltonian separated by physical, Minkowski time evolution. For physical quark masses, the kaon mass lies above the two- and three-pion thresholds and, as a result, this long-range matrix element receives contributions from intermediate on-shell $2\pi$ and $3\pi$ states. These contributions cannot easily be captured in a finite Euclidean spacetime, meaning that such matrix elements are not directly accessible via lattice QCD. In this talk, we present a strategy for combining quantities that can be extracted in numerical lattice QCD calculations in order to reproduce the physical, infinite-volume long-range amplitude for $K^0-\bar{K}^0$. The key novelty relative to published work is that we fully include the effects of three-particle states that were previously neglected. The strategy is built on existing formalism for long-range matrix elements with two-particle intermediate states, together with the relativistic-field-theory finite-volume formalism for extracting three-hadron weak decays.}

\FullConference{%
39th International Symposium on Lattice Field Theory - Lattice2022 \\
8-13 August 2022 \\
Bonn, Germany
}

\begin{document}
\maketitle

\section{Introduction}

Neutral kaon oscillations provide one of many promising avenues for constraining (or discovering) new physics beyond the Standard Model. But such constraints (or discoveries) are only possible if the Standard Model prediction is sufficiently reliable. In this spirit, the RBC/UKQCD collaboration has made significant progress in providing first-principles lattice-QCD determinations of the mass splitting, $\Delta M$, of the two neutral-kaon mass eigenstates ($K_L - K_S$), and the closely related CP violating parameter $\epsilon_K$~\cite{Christ:2012se,Bai:2014cva,Christ:2014qwa,Bai:2018mdv}. The mass splitting is determined from a long-range matrix element of two weak-Hamiltonian densities between a kaon and an anti-kaon.

Motivated by such ongoing calculations, this proceedings and the preprint (to appear) are dedicated to the formal challenge of relating finite-volume Euclidean spacetime correlation functions to physical, infinite-volume Minkowksi-signature matrix elements, in particular when three-particle effects play a role. Such relations are of direct relevance to lattice QCD calculations, as these necessarily estimate finite-volume Euclidean-signature correlation functions that must be related to the physical observable.

Once the external kaon states are created, intermediate multi-particle states can lead to exponentials that grow as a function of the Euclidean time separation between the two currents. After such growing exponentials are removed, the resulting estimator has poles arising from the discrete finite-volume spectrum as well as power-like finite-volume artefacts. These are related to finite-volume effects that arise in other multi-hadron observables, such as $\pi \pi \to \pi \pi$ scattering and $K \to 2\pi$ decays. It is well-known how to treat these finite-volume effects for the determinations of two-body scattering amplitudes using the methods first introduced by L\"uscher~\cite{Rummukainen:1995vs,Kim:2005gf,Fu:2011xz,Briceno:2012yi,Hansen:2012tf,Feng:2004ua,Briceno:2014oea} and for decay and transition amplitudes using the formalism of Lellouch and L{\"u}scher and subsequent extensions~\cite{Lellouch:2000pv,Briceno:2014uqa,Briceno:2015csa, Briceno:2021xlc, Meyer:2011um, Feng:2014gba}.

In addition to $2\pi$ effects, $3\pi$ physics could play an important role in quantifying CP violating observables from neutral kaon oscillations. This is because for the longer-lived mass eigenstate ($K_L$), the decay $K_L \to 3\pi$ is the dominant mode. At physical quark masses, $3\pi$ dynamics arising from $K\to 3\pi$ decays also play an important role in computing long-range matrix elements from QCD. Three-pion states induce similar finite-volume effects as those in the two-pion sector, and extensions to the two-body formalism have been developed to constrain three-body scattering amplitudes from discrete finite-volume spectra~\cite{Hansen:2014eka, Hansen:2015zga, Hansen:2020zhy}, with first applications of the formalism published in refs.~\cite{Hansen:2020otl, Fischer:2020jzp,Culver:2019vvu,Horz:2019rrn}. Recently, the framework has been extended to extract three-body decay and transition amplitudes~\cite{Hansen:2021ofl}.

In these proceedings, we present a new formalism which builds upon these previously derived relations to compute long-range matrix elements including both $2\pi$ and $3\pi$ dynamics from finite-volume Euclidean correlation functions. This work can find immediate applications to ongoing calculations of $K^0 - \bar{K}^0$ oscillations~\cite{Christ:2012se,Bai:2014cva,Christ:2014qwa,Bai:2018mdv}, to ensure the proper systematic removal of all finite-volume and Euclidean artefacts.

\section{The $K^0\leftrightarrow \bar{K}^0$ physical matrix element from Euclidean correlators}

As recently reviewed in refs.~\cite{Christ:2015pwa,Christ:2012se}, $K^0-\bar{K}^0$ mixing is parameterized at next-to-leading order in the weak interaction by the mass splitting $\Delta M$. This splitting is typically written in terms of the mass matrix $M_{ab}$, which couples different mass eigenstates, by the relation $\Delta M = 2 M_{0\bar{0}}$.

It is well known that the mass splitting can be determined from the principal value contribution of an infinite-volume Minwkowki correlation function,
\begin{align}
\Delta M &= \frac{1}{2 m_K} {\rm Re}   [\mathcal{T}]
\label{eq:DeltaM}\\
&\equiv \frac{1}{2 m_K}
{\rm Re}\,\,i\!\int\! {\rm d}^4{x}_1
e^{-\epsilon  |x_1^0 |}
\langle {\bar{K}^0},\textbf{0};\infty| \mathrm{T}_{\sf M} \{ \mathcal{H}_{W\!,{\sf M}} (x_1) \mathcal{H}_W (0) \} |{K^0},\textbf{0};\infty\rangle
\bigg|_{\epsilon=0},
\label{eq:iT}
\end{align}
where $\mathcal{H}_{W\!,{\sf M}}(x_1)$ is the weak-Hamiltonian density evaluated at the $x_1$ spacetime point, $\mathrm{T}_{\sf M}$ is the time-ordering operator, the ${\sf M}$ subscript emphasizes Minkowski time dependence, and the infinite-volume kaon states have the standard relativistic normalization, i.e., $\langle {K^0},\textbf{P}_f;\infty|{K^0},\textbf{P}_i;\infty\rangle
=2 \omega_{K,\textbf P_i}  \,(2\pi)^3 \delta^3(\textbf{P}_f-\textbf{P}_i)$, where $\omega_{K,\textbf P} = \sqrt{m_K^2 + \textbf P^2}$ and $m_K$ is the physical kaon mass. The $\epsilon$-dependent exponential enforces the $i \epsilon$ prescription defining $\mathcal T$.

\begin{figure}[t]
	\begin{center}
	\includegraphics[width=.95\textwidth]{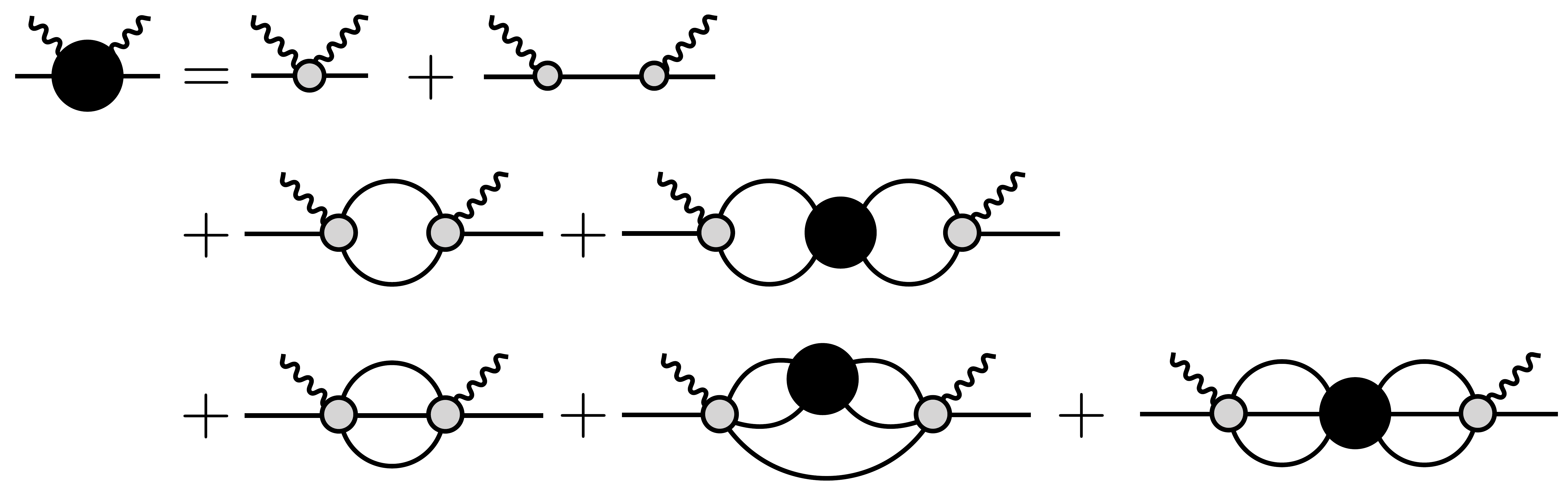}
	\caption{Shown is the $i\mathcal{T}$ amplitude, defined in eq.~\eqref{eq:iT}. The grey circles represent kernels coupling the various hadronic legs to one- and two-insertions of the weak Hamiltonian. The black circles represent purely hadronic two- and three-particle scattering amplitudes, defined diagrammatically in figure~\ref{fig:iM}.}
	\label{fig:iT}
	\end{center}
\end{figure}

In what follows, we make use of diagrammatic representations of correlation functions in finite  and infinite volumes. The one introduced above, $\mathcal{T}$, can be defined diagrammatically using now-standard techniques introduced in refs.~\cite{Briceno:2019opb,Briceno:2022omu, Briceno:2020vgp,Sherman:2022tco}. In figure~\ref{fig:iT} we show $i\mathcal{T}$, where in the Feynman diagrams we represent the insertion of the weak Hamiltonian multiplied by a factor of $i$ with a wiggly line. This explains the overall factor of $(-1)$ relating $\mathcal{T}$ and $\Delta M$ in eq.~\eqref{eq:DeltaM}.

In figure~\ref{fig:iT}, the external legs represent physical kaons, which are amputated and placed on shell. The internal legs represent pion propagators, whose momenta are in general off-shell. The grey circles represent short-distance kernels, while the black circles are off-shell extensions of the purely hadronic two- and three-particle scattering amplitudes. The key feature needed in the definition of every kernel is that they are analytic functions for energies in the vicinity of the kaon mass, i.e. their singularities are far from the kinematic region of interest.

Applying the same techniques used in previous works in the two- and three-particle sector~\cite{Briceno:2019opb,Briceno:2022omu, Briceno:2020vgp,Sherman:2022tco, Hansen:2021ofl}, one can isolate the singularities of the infinite-volume, physical amplitude and write it in terms purely on-shell quantities that are singularity free in the kinematic region of interest. This is, however, unnecessary for our immediate goal, which is to find a relationship between $\Delta M$ and correlation functions that be accessible via lattice QCD.

\begin{figure}[t]
	\begin{center}
	\includegraphics[width=.95\textwidth]{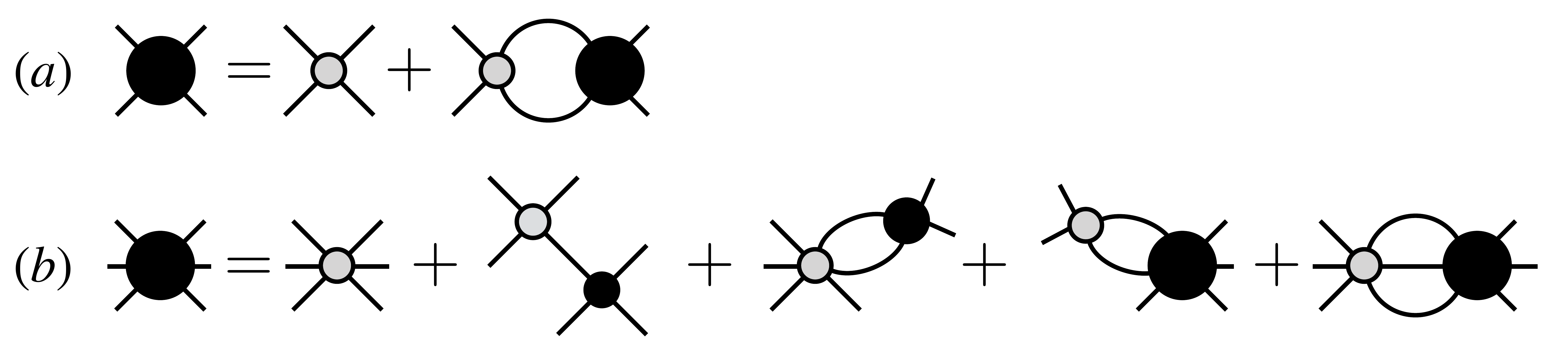}
	\caption{Shown are the purely hadronic (a) two- and (b) three-particle scattering amplitudes. The grey circles represent hadronic kernels.}
	\label{fig:iM}
	\end{center}
\end{figure}

In a finite volume with Euclidean time signature, as is in numerical lattice QCD calculations, integrals such as eq.~\eqref{eq:iT} are not directly accessible. Instead, one can constrain the analogous finite-volume Euclidean correlation function
\begin{align}
    C_L(\tau)
    \equiv
    2m_KL^3\int_L\! {\rm d}^3 \textbf{x}_1
   \,  \langle {\bar{K}^0}, \textbf{0};L| \mathrm{T}_{\sf E} \{ \mathcal{H}_{W \!,{\sf E}} (\tau, \textbf{x}_1) \mathcal{H}_W (0) \} |{K^0}, \textbf{0};L\rangle \, ,
    \label{eq:CLtau}
    \end{align}
where $\tau$ is the Euclidean time and as is emphasized by the ${\sf E}$ subscript. Another important distinction is that the finite-volume states have been normalized to unity, and the multiplicative factor of $2m_K L^3$ is sufficient for local matrix elements of single-particle states to be exponentially close to their infinite-volume ones~\cite{Briceno:2014uqa}. For non-local matrix elements, like the one we are after here, one must provide a more careful treatment as discussed below.

Na\"ively, one might hope that the Euclidean time integral of $C_L(\tau)$ could be related to $\Delta M$, since it looks superficially like the analytical continuation of the correct Minkowski-signature Fourier transform. However, since the physical kaon mass satisfies $m_K>3 m_\pi$, the integral of $ C_L(\tau)$ is not convergent due to finite-volume two- and three-pion states.\footnote{We comment that, in a practical lattice calculation, one could break up the weak Hamiltonian into separate operators that couple either to two- or three-pion states only. In this way the analysis of growing exponentials and finite-volume effects could be decoupled in the two sectors.} This is easily seen if one performs a spectral decomposition of eq.~\eqref{eq:CLtau},
\begin{align}
C_L(\tau)
=
\sum_{n}
{c_n} \left[e^{-(E_n - m_k) \tau}\Theta(\tau)
+e^{(E_n - m_k) \tau}\Theta(-\tau)\right],
\end{align}
where $c_n$ is defined in terms of local finite-volume matrix elements of the weak Hamiltonian,
\begin{align}
c_n
\equiv 2 m_K L^6   \, \langle \bar{K}^0,\textbf{0};L|  \mathcal{H}_{W} (0)
|
n,\textbf{0};L\rangle \,
\langle n,\textbf{0};L
|\mathcal{H}_{W} (0)   |{K}^0,\textbf{0};L\rangle.
\end{align}
One observes that the resulting integral is not convergent because a finite set of intermediate states would satisfy $E_n > m_K$. Once again, we have normalized the intermediate finite-volume states to unity regardless of whether they are associated with one-, two-, or three-particle states. The integral can be rendered finite by subtracting a finite number of divergent terms. We define the subtracted correlator,
\begin{align}
 C_L^{ >}(\tau)
\equiv
C_L (\tau)-\sum_{n=0}^N c_n
\left[e^{-(E_n - m_k) \tau}\Theta(\tau)
+e^{(E_n - m_k) \tau}\Theta(-\tau)\right],
\end{align}
where the sum on $n$ is over the minimum number of states $N$ needed to make the $\tau$ integral finite.

The contributions from the first $N$ terms in the series can be evaluated directly in a finite-Minkowski spacetime by performing the integral over the time-dependent Minkowski correlator associated with these finite number of states,
\begin{align}
\mathcal{T}_L^{<}
&=i
\sum_{n=0}^N
c_n
\,
\int_{-\infty}^\infty
{\rm d}x_1^0 e^{-\epsilon \left|x_1^0\right|}
\left[e^{-i(E_n - m_k) x_1^0}
\Theta(x_1^0)
+e^{i(E_n - m_k) x_1^0}
\Theta(-x_1^0)\right]_{\epsilon =0} \,,
\\
&=
\sum_{n=0}^N
\frac{2c_n}{E_n-m_K}.
\end{align}
It is worth emphasizing that this integral does not need to evaluated numerically. Instead, all that is needed to evaluate this contribution is the finite-volume spectra and matrix elements of low-lying states.

\begin{figure}[t]
	\begin{center}
	\includegraphics[width=.95\textwidth]{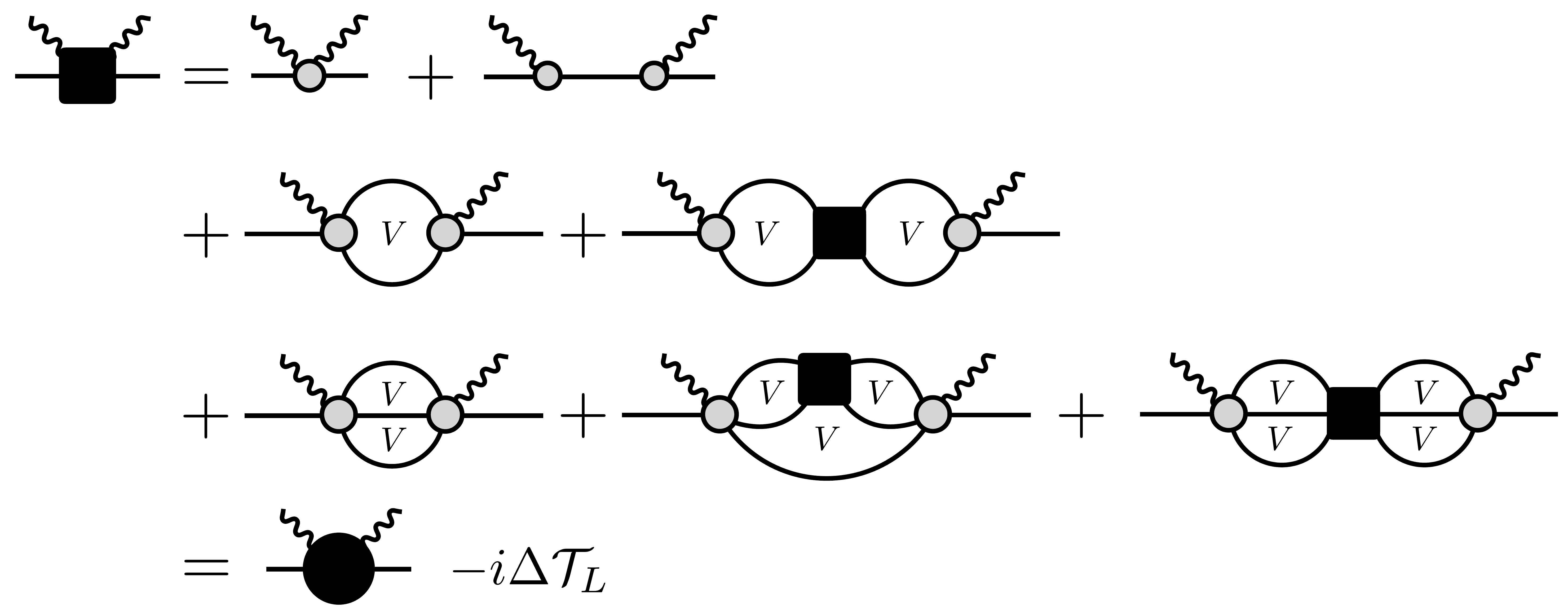}
	\caption{Shown in the finite-volume version of $iT$, defined in figure~\ref{fig:iT}. The building blocks are the same, except the loops are defined in terms of discrete momenta. The black squares are finite-volume extensions of amplitudes defined in figure~\ref{fig:iML}.
	}
	\label{fig:iTL}
	\end{center}
\end{figure}

This contribution is necessary to define a finite-volume Minkowski-signature estimator of $\Delta M$. We do this by recognizing that the combination of the two previously identified terms gives us a definition of the estimator, $\mathcal{T}_L$, according to
\begin{align}
\mathcal{T}_L =
\mathcal{T}^{<}_L
+\int \! {\rm d}\tau \,
C^>_L(\tau).
\label{eq:TLv1}
\end{align}
It is this quantity that suffers from power-law finite-volume artefacts, as well as finite-volume dependent poles. Both must be removed to reach a reliable calculation of the infinite-volume observable.

To remove volume effects, the estimator $\mathcal T_L$ can be studied diagrammatically to all-orders. As shown in figure~\ref{fig:iTL}, this is defined in the same exact way as $\mathcal{T}$ with the loops integrated over continuous momenta replaced by sums over discrete momenta. One can write each finite-volume loop as an infinite-volume loop followed by the difference between the two. For two- and three-particle loops, this difference leads to power-law finite-volume artefacts that can be written in terms of purely on-shell amplitudes and known finite-volume functions. In general, one finds that
\begin{align}
\mathcal{T}_L =
\mathcal{T} - \Delta \mathcal{T}_L,
\label{eq:TLv2}
\end{align}
where below we give an exact form for $\Delta \mathcal{T}_L$. Combining eqs.~\eqref{eq:TLv1} and~\eqref{eq:TLv2}, one finds that the desired amplitude, $\mathcal{T}$, can be reconstructed from different known quantities using the identity,
\begin{align}
\mathcal{T}
=
\left[\mathcal{T}_L^{
<}
+
\Delta \mathcal{T}_{L}
\right]
+
\int \! {\rm d}\tau \,
C^>_L(\tau).
\label{eq:result}
\end{align}
%

\begin{figure}[t]
	\begin{center}
	\includegraphics[width=.95\textwidth]{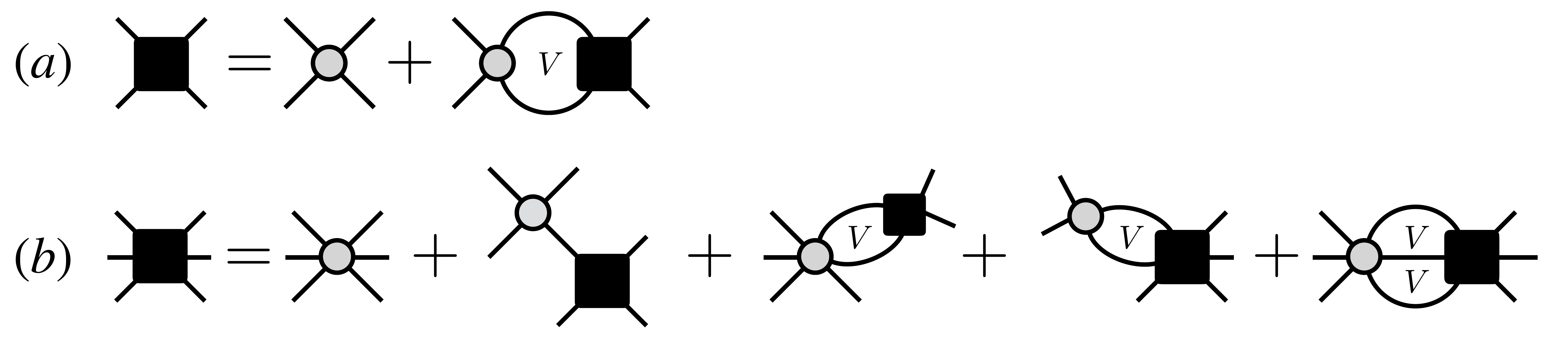}
	\caption{Shown are the finite-volume version of the (a) two- and (b) three-body scattering amplitudes defined in figure~\ref{fig:iM}. The kernels are the same as appear in figure~\ref{fig:iM}. The label $V$ inside the loops are emphasizing the fact that these are defined in a finite volume. }
	\label{fig:iML}
	\end{center}
\end{figure}

\subsection{The finite-volume correction}

Here we provide a sketch of the derivation of eq.~\eqref{eq:TLv2} and give the explicit form of $\Delta T_L$, given in eq.~\eqref{eq:correction} below.\footnote{A detailed derivation will await a future publication.} In what follows, we rely on the diagrammatic representation shown in figure~\ref{fig:iTL}. This is defined in terms of kernels (grey circles) and finite-volume amplitudes (black squares). The off-shell extension of finite-volume hadronic amplitudes are defined in figure~\ref{fig:iML}. We assume that the volumes are sufficiently large, $m_\pi L\gg1$, such that exponentially suppressed effects of the order $\mathcal{O}(e^{-m_\pi L})$ are a subdominant effect in the numerical lattice calculations in which these formulas would be applied, and can be safely ignored. This means that the masses and kernels used in the definition of figure~\ref{fig:iTL} are safely approximated to be their infinite-volume counterparts.

We break down the derivation of $\Delta \mathcal{T}_L$ in three parts, defined by the three lines of the right-hand side of figure~\ref{fig:iTL}. The first line includes possible short distance contributions where no intermediate states may go on-shell as well as the single particle pole. The latter can be due to an intermediate single pion propagator. Because the volumes considered satisfy $m_\pi L\gg 1$, this contribution which we label as $\mathcal{T}_L^{(1)}$ is equal to the infinite-volume analogue up to exponentially suppressed corrections,
\begin{align}
i\mathcal{T}_L^{(1)}
=
i\mathcal{T}^{(1)}+\mathcal{O}(e^{-m_\pi L}),
\label{eq:iTL1}
\end{align}
where, for simplicity, we leave the dependence on kinematics implicit.

The next contribution is shown in the second line of figure~\ref{fig:iTL}, which includes contributions associated with possible two-particle intermediate states. These diagrams can have power-law finite effects associated with these particles going on shell. It is well known that these contributions can be parameterized by a geometric function, typically labelled $F$, and the infinite-volume two-particle scattering amplitude, $\mathcal{M}_2$~\cite{Briceno:2019opb,Briceno:2022omu}. In short, we find that piece of the finite-volume amplitude, $\mathcal{T}_L^{(2)}$, can be written in terms of a corresponding infinite-volume term, $\mathcal{T}^{(2)}$, as
\begin{align}
i\mathcal{T}_L^{(2)}
=
i\mathcal{T}^{(2)}
+i\mathcal{H}_2\, \frac{i}{F^{-1}+\mathcal{M}_2}\,
i\mathcal{H}_2
+\mathcal{O}(e^{-m_\pi L}),
\label{eq:iTL2}
\end{align}
where $\mathcal{H}_2$ is the $K\to \pi\pi$ amplitude.
In addition to not explicitly showing the dependence on the kinematics, we also leave the volume dependence implicit. It is well known that this amplitude can be obtained from finite-volume matrix elements~\cite{Lellouch:2000pv,Briceno:2014uqa} and lattice QCD calculations have been under way~\cite{Blum:2011ng,Blum:2011pu,Blum:2012uk,Blum:2015ywa}. This correction was the main focus of Ref.~\cite{Christ:2015pwa, Briceno:2019opb} and is the only one required for quark masses where $2m_\pi < m_K < 3m_\pi$. For such quark masses, the intermediate three-particle states shown explicitly in the third line of figure~\ref{fig:iTL} can not go on shell. As a result, these will have finite-volume effects that are exponentially suppressed and fall under the class of contributions defined by eq.~\eqref{eq:iTL1}. For physical quark masses, $ 3m_\pi <m_K$, and as a result we must treat these diagrams carefully.

For the evaluation of the diagrams shown in the third line of figure~\ref{fig:iTL}, we make use of the existing three-body finite-volume formalism~\cite{Hansen:2014eka,Hansen:2015zga,Hansen:2021ofl}. In fact, one can just read off the answer from the second term in eq.~(42) if Ref.~\cite{Hansen:2014eka},\footnote{The one modification needed for adopting this result is $A\to i\mathbf{H}_3$, due to the fact that the interpolators in the aforementioned reference do not include factors of $i$.}
\begin{align}
i\mathcal{T}_L^{(3)}
=
i\mathcal{T}^{(3)}
+i\mathbf{H}_3\, \frac{i}{F^{-1}_3+\mathcal{K}_{\rm df,3}}\,
i\mathbf{H}_3
+\mathcal{O}(e^{-m_\pi L}),
\label{eq:iTL3}
\end{align}
where $F_3$ is a known finite-volume function that depends on the two-body scattering amplitude~\cite{Hansen:2014eka}, $\mathcal{K}_{\rm df,3}$ is the $3\pi$ K matrix that can be constrained from the $3\pi$ spectrum and directly related to the $3\pi\to3\pi$ scattering amplitude~\cite{Hansen:2014eka, Hansen:2015zga}, and $\mathbf{H}_3$ is proportional to the $K\to 3\pi$ decay amplitude that can be obtained from the corresponding finite-volume matrix element~\cite{Hansen:2021ofl}.

Adding eqs.~\eqref{eq:iTL1}, \eqref{eq:iTL2} and \eqref{eq:iTL3} and equating them eq.~\eqref{eq:TLv2}, we find an explicit expression for the correction term,
\begin{align}
\Delta \mathcal{T}_L=
\mathcal{H}_2\, \frac{1}{F^{-1}+\mathcal{M}_2}\,
\mathcal{H}_2
+\mathbf{H}_3\, \frac{1}{F^{-1}_3+\mathcal{K}_{\rm df,3}}\,
\mathbf{H}_3+\mathcal{O}(e^{-m_\pi L}).
\label{eq:correction}
\end{align}
It is worth emphasizing that all of these components can be obtained from the finite-volume $2\pi$, $3\pi$ spectra and the $K\to 2\pi$, $K\to 3\pi$ weak matrix elements.

\section*{Conclusion and Outlook}

We have laid out a complete framework to compute the $K_S-K_L$ mass splitting from lattice QCD for physical kaon mass, which includes physical effects from $2\pi$ and $3\pi$ decay dynamics. Our formalism relies on a two-step procedure to obtain the infinite-volume $K^0-\bar{K}^0$ mixing Minkowski matrix elements from finite-volume Euclidean lattices. First, we relate compute finite-volume Euclidean-signature matrix elements from Lattice QCD, and use the result eq.~\eqref{eq:TLv1} in order to obtain the finite-volume amplitude $\mathcal{T}_L$. Second, We remove the finite-volume effects which are power-law in the volume $L$ with eq.~\eqref{eq:TLv2}, where the correction $\Delta \mathcal{T}_L$, eq.~\eqref{eq:correction}, is given in terms of previously defined on-shell quantities which can be obtained with various finite-volume formalisms for the $2\pi$ and $3\pi$ scattering amplitudes as well as the $K\to \pi\pi$ and $K\to 3\pi$ decay amplitudes. This complete procedure, given in eq.~\eqref{eq:result}, allows us to determine the infinite-volume Minkowski signature amplitude $\mathcal{T}$, and in turn determine the mass splitting $\Delta M = {\rm Re} [\mathcal{T}] / (2m_K)$.

\section*{Acknowledgments}

RAB and AWJ are supported in part by USDOE grant No. DE-AC05-06OR23177, under which Jefferson Science Associates, LLC, manages and operates Jefferson Lab. This work is also supported from the USDOE Early Career award, contract de-sc0019229. MTH is supported by UKRI Future Leader Fellowship MR/T019956/1 and in part by UK STFC grant ST/P000630/1.

\bibliographystyle{JHEP}
\bibliography{refs}

\providecommand{\href}[2]{#2}\begingroup\raggedright\begin{thebibliography}{10}

\bibitem{Christ:2012se}
\textsc{(RBC, UKQCD)}, N.~H. Christ, T.~Izubuchi, C.~T. Sachrajda, A.~Soni and
  J.~Yu, \emph{{Long distance contribution to the KL-KS mass difference}},
  \href{http://dx.doi.org/10.1103/PhysRevD.88.014508}{\emph{Phys. Rev. D} {\bf
  88} (2013) 014508}, [\href{http://arxiv.org/abs/1212.5931}{{\tt 1212.5931}}].

\bibitem{Bai:2014cva}
Z.~Bai, N.~H. Christ, T.~Izubuchi, C.~T. Sachrajda, A.~Soni and J.~Yu,
  \emph{{$K_L-K_S$ Mass Difference from Lattice QCD}},
  \href{http://dx.doi.org/10.1103/PhysRevLett.113.112003}{\emph{Phys. Rev.
  Lett.} {\bf 113} (2014) 112003}, [\href{http://arxiv.org/abs/1406.0916}{{\tt
  1406.0916}}].

\bibitem{Christ:2014qwa}
\textsc{(RBC, UKQCD)}, N.~Christ, T.~Izubuchi, C.~T. Sachrajda, A.~Soni and
  J.~Yu, \emph{{Calculating the $K_L-K_S$ mass difference and $\epsilon_K$ to
  sub-percent accuracy}},
  \href{http://dx.doi.org/10.22323/1.187.0397}{\emph{PoS} {\bf LATTICE2013}
  (2014) 397}, [\href{http://arxiv.org/abs/1402.2577}{{\tt 1402.2577}}].

\bibitem{Bai:2018mdv}
Z.~Bai, N.~H. Christ and C.~T. Sachrajda, \emph{{The $K_L$ - $K_S$ Mass
  Difference}}, \href{http://dx.doi.org/10.1051/epjconf/201817513017}{\emph{EPJ
  Web Conf.} {\bf 175} (2018) 13017}.

\bibitem{Rummukainen:1995vs}
K.~Rummukainen and S.~A. Gottlieb, \emph{Resonance scattering phase shifts on a
  nonrest frame lattice},
  \href{http://dx.doi.org/10.1016/0550-3213(95)00313-H}{\emph{Nucl. Phys.} {\bf
  B450} (1995) 397--436}, [\href{http://arxiv.org/abs/hep-lat/9503028}{{\tt
  hep-lat/9503028}}].

\bibitem{Kim:2005gf}
C.~Kim, C.~Sachrajda and S.~R. Sharpe, \emph{Finite-volume effects for
  two-hadron states in moving frames},
  \href{http://dx.doi.org/10.1016/j.nuclphysb.2005.08.029}{\emph{Nucl.Phys.}
  {\bf B727} (2005) 218--243},
  [\href{http://arxiv.org/abs/hep-lat/0507006}{{\tt hep-lat/0507006}}].

\bibitem{Fu:2011xz}
Z.~Fu, \emph{{Rummukainen-Gottlieb's formula on two-particle system with
  different mass}},
  \href{http://dx.doi.org/10.1103/PhysRevD.85.014506}{\emph{Phys. Rev.} {\bf
  D85} (2012) 014506}, [\href{http://arxiv.org/abs/1110.0319}{{\tt
  1110.0319}}].

\bibitem{Briceno:2012yi}
R.~A. Brice\~no and Z.~Davoudi, \emph{Moving multichannel systems in a finite
  volume with application to proton-proton fusion},
  \href{http://dx.doi.org/10.1103/PhysRevD.88.094507}{\emph{Phys. Rev. D. 88,}
  {\bf 094507} (2013) 094507}, [\href{http://arxiv.org/abs/1204.1110}{{\tt
  1204.1110}}].

\bibitem{Hansen:2012tf}
M.~T. Hansen and S.~R. Sharpe, \emph{Multiple-channel generalization of
  lellouch-luscher formula},
  \href{http://dx.doi.org/10.1103/PhysRevD.86.016007}{\emph{Phys.Rev.} {\bf
  D86} (2012) 016007}, [\href{http://arxiv.org/abs/1204.0826}{{\tt
  1204.0826}}].

\bibitem{Feng:2004ua}
X.~Feng, X.~Li and C.~Liu, \emph{{Two particle states in an asymmetric box and
  the elastic scattering phases}},
  \href{http://dx.doi.org/10.1103/PhysRevD.70.014505}{\emph{Phys. Rev.} {\bf
  D70} (2004) 014505}, [\href{http://arxiv.org/abs/hep-lat/0404001}{{\tt
  hep-lat/0404001}}].

\bibitem{Briceno:2014oea}
R.~A. Brice\~no, \emph{Two-particle multichannel systems in a finite volume
  with arbitrary spin},
  \href{http://dx.doi.org/10.1103/PhysRevD.89.074507}{\emph{Phys.Rev.} {\bf
  D89} (2014) 074507}, [\href{http://arxiv.org/abs/1401.3312}{{\tt
  1401.3312}}].

\bibitem{Lellouch:2000pv}
L.~Lellouch and M.~Luscher, \emph{Weak transition matrix elements from finite
  volume correlation functions}, {\emph{Commun.Math.Phys.} {\bf 219} (2001)
  31--44}, [\href{http://arxiv.org/abs/hep-lat/0003023}{{\tt
  hep-lat/0003023}}].

\bibitem{Briceno:2014uqa}
R.~A. Brice\~no, M.~T. Hansen and A.~Walker-Loud, \emph{{Multichannel 1
  $\rightarrow$ 2 transition amplitudes in a finite volume}},
  \href{http://dx.doi.org/10.1103/PhysRevD.91.034501}{\emph{Phys. Rev. D} {\bf
  91} (2015) 034501}, [\href{http://arxiv.org/abs/1406.5965}{{\tt 1406.5965}}].

\bibitem{Briceno:2015csa}
R.~A. Brice\~no and M.~T. Hansen, \emph{{Multichannel 0 $\to$ 2 and 1 $\to$ 2
  transition amplitudes for arbitrary spin particles in a finite volume}},
  \href{http://dx.doi.org/10.1103/PhysRevD.92.074509}{\emph{Phys. Rev. D} {\bf
  92} (2015) 074509}, [\href{http://arxiv.org/abs/1502.04314}{{\tt
  1502.04314}}].

\bibitem{Briceno:2021xlc}
R.~A. Brice\~no, J.~J. Dudek and L.~Leskovec, \emph{{Constraining
  $1+\mathcal{J}\to 2$ coupled-channel amplitudes in finite-volume}},
  \href{http://dx.doi.org/10.1103/PhysRevD.104.054509}{\emph{Phys. Rev. D} {\bf
  104} (2021) 054509}, [\href{http://arxiv.org/abs/2105.02017}{{\tt
  2105.02017}}].

\bibitem{Meyer:2011um}
H.~B. Meyer, \emph{Lattice qcd and the timelike pion form factor},
  \href{http://dx.doi.org/10.1103/PhysRevLett.107.072002}{\emph{Phys. Rev.
  Lett.} {\bf 107} (2011) 072002}, [\href{http://arxiv.org/abs/1105.1892}{{\tt
  1105.1892}}].

\bibitem{Feng:2014gba}
X.~Feng, S.~Aoki, S.~Hashimoto and T.~Kaneko, \emph{Timelike pion form factor
  in lattice qcd},
  \href{http://dx.doi.org/10.1103/PhysRevD.91.054504}{\emph{Phys. Rev.} {\bf
  D91} (2015) 054504}, [\href{http://arxiv.org/abs/1412.6319}{{\tt
  1412.6319}}].

\bibitem{Hansen:2014eka}
M.~T. Hansen and S.~R. Sharpe, \emph{{Relativistic, model-independent,
  three-particle quantization condition}},
  \href{http://dx.doi.org/10.1103/PhysRevD.90.116003}{\emph{Phys. Rev. D} {\bf
  90} (2014) 116003}, [\href{http://arxiv.org/abs/1408.5933}{{\tt 1408.5933}}].

\bibitem{Hansen:2015zga}
M.~T. Hansen and S.~R. Sharpe, \emph{{Expressing the three-particle
  finite-volume spectrum in terms of the three-to-three scattering amplitude}},
  \href{http://dx.doi.org/10.1103/PhysRevD.92.114509}{\emph{Phys. Rev. D} {\bf
  92} (2015) 114509}, [\href{http://arxiv.org/abs/1504.04248}{{\tt
  1504.04248}}].

\bibitem{Hansen:2020zhy}
M.~T. Hansen, F.~Romero-L\'opez and S.~R. Sharpe, \emph{{Generalizing the
  relativistic quantization condition to include all three-pion isospin
  channels}}, \href{http://dx.doi.org/10.1007/JHEP07(2020)047}{\emph{JHEP} {\bf
  07} (2020) 047}, [\href{http://arxiv.org/abs/2003.10974}{{\tt 2003.10974}}].
  [Erratum: JHEP 02, 014 (2021)].

\bibitem{Hansen:2020otl}
\textsc{(Hadron Spectrum)}, M.~T. Hansen, R.~A. Brice\~no, R.~G. Edwards, C.~E.
  Thomas and D.~J. Wilson, \emph{{Energy-Dependent $\pi^+ \pi^+ \pi^+$
  Scattering Amplitude from QCD}},
  \href{http://dx.doi.org/10.1103/PhysRevLett.126.012001}{\emph{Phys. Rev.
  Lett.} {\bf 126} (2021) 012001}, [\href{http://arxiv.org/abs/2009.04931}{{\tt
  2009.04931}}].

\bibitem{Fischer:2020jzp}
M.~Fischer, B.~Kostrzewa, L.~Liu, F.~Romero-L\'opez, M.~Ueding and C.~Urbach,
  \emph{{Scattering of two and three physical pions at maximal isospin from
  lattice QCD}},
  \href{http://dx.doi.org/10.1140/epjc/s10052-021-09206-5}{\emph{Eur. Phys. J.
  C} {\bf 81} (2021) 436}, [\href{http://arxiv.org/abs/2008.03035}{{\tt
  2008.03035}}].

\bibitem{Culver:2019vvu}
C.~Culver, M.~Mai, R.~Brett, A.~Alexandru and M.~D\"oring, \emph{{Three pion
  spectrum in the $I=3$ channel from lattice QCD}},
  \href{http://dx.doi.org/10.1103/PhysRevD.101.114507}{\emph{Phys. Rev. D} {\bf
  101} (2020) 114507}, [\href{http://arxiv.org/abs/1911.09047}{{\tt
  1911.09047}}].

\bibitem{Horz:2019rrn}
B.~H\"orz and A.~Hanlon, \emph{{Two- and three-pion finite-volume spectra at
  maximal isospin from lattice QCD}},
  \href{http://dx.doi.org/10.1103/PhysRevLett.123.142002}{\emph{Phys. Rev.
  Lett.} {\bf 123} (2019) 142002}, [\href{http://arxiv.org/abs/1905.04277}{{\tt
  1905.04277}}].

\bibitem{Hansen:2021ofl}
M.~T. Hansen, F.~Romero-L\'opez and S.~R. Sharpe, \emph{{Decay amplitudes to
  three hadrons from finite-volume matrix elements}},
  \href{http://dx.doi.org/10.1007/JHEP04(2021)113}{\emph{JHEP} {\bf 04} (2021)
  113}, [\href{http://arxiv.org/abs/2101.10246}{{\tt 2101.10246}}].

\bibitem{Christ:2015pwa}
N.~H. Christ, X.~Feng, G.~Martinelli and C.~T. Sachrajda, \emph{{Effects of
  finite volume on the $K_L$-$K_S$ mass difference}},
  \href{http://dx.doi.org/10.1103/PhysRevD.91.114510}{\emph{Phys. Rev. D} {\bf
  91} (2015) 114510}, [\href{http://arxiv.org/abs/1504.01170}{{\tt
  1504.01170}}].

\bibitem{Briceno:2019opb}
R.~A. Brice\~no, Z.~Davoudi, M.~T. Hansen, M.~R. Schindler and A.~Baroni,
  \emph{{Long-range electroweak amplitudes of single hadrons from Euclidean
  finite-volume correlation functions}},
  \href{http://dx.doi.org/10.1103/PhysRevD.101.014509}{\emph{Phys. Rev. D} {\bf
  101} (2020) 014509}, [\href{http://arxiv.org/abs/1911.04036}{{\tt
  1911.04036}}].

\bibitem{Briceno:2022omu}
R.~A. Brice\~no, A.~W. Jackura, A.~Rodas and J.~V. Guerrero, \emph{{Prospects
  for $\gamma^\star \gamma^\star \to \pi \pi$ via lattice QCD}},
  \href{http://arxiv.org/abs/2210.08051}{{\tt 2210.08051}}.

\bibitem{Briceno:2020vgp}
R.~A. Brice\~no, A.~W. Jackura, F.~G. Ortega-Gama and K.~H. Sherman,
  \emph{{On-shell representations of two-body transition amplitudes: Single
  external current}},
  \href{http://dx.doi.org/10.1103/PhysRevD.103.114512}{\emph{Phys. Rev. D} {\bf
  103} (2021) 114512}, [\href{http://arxiv.org/abs/2012.13338}{{\tt
  2012.13338}}].

\bibitem{Sherman:2022tco}
K.~H. Sherman, F.~G. Ortega-Gama, R.~A. Brice\~no and A.~W. Jackura,
  \emph{{Two-current transition amplitudes with two-body final states}},
  \href{http://dx.doi.org/10.1103/PhysRevD.105.114510}{\emph{Phys. Rev. D} {\bf
  105} (2022) 114510}, [\href{http://arxiv.org/abs/2202.02284}{{\tt
  2202.02284}}].

\bibitem{Blum:2011ng}
T.~Blum et~al., \emph{{The $K\to(\pi\pi)_{I=2}$ Decay Amplitude from Lattice
  QCD}}, \href{http://dx.doi.org/10.1103/PhysRevLett.108.141601}{\emph{Phys.
  Rev. Lett.} {\bf 108} (2012) 141601},
  [\href{http://arxiv.org/abs/1111.1699}{{\tt 1111.1699}}].

\bibitem{Blum:2011pu}
T.~Blum et~al., \emph{{$K$ to $\pi\pi$ Decay amplitudes from Lattice QCD}},
  \href{http://dx.doi.org/10.1103/PhysRevD.84.114503}{\emph{Phys. Rev. D} {\bf
  84} (2011) 114503}, [\href{http://arxiv.org/abs/1106.2714}{{\tt 1106.2714}}].

\bibitem{Blum:2012uk}
T.~Blum et~al., \emph{{Lattice determination of the $K \to (\pi\pi)_{I=2}$
  Decay Amplitude $A_2$}},
  \href{http://dx.doi.org/10.1103/PhysRevD.86.074513}{\emph{Phys. Rev. D} {\bf
  86} (2012) 074513}, [\href{http://arxiv.org/abs/1206.5142}{{\tt 1206.5142}}].

\bibitem{Blum:2015ywa}
T.~Blum et~al., \emph{{$K \rightarrow \pi\pi$ $\Delta I=3/2$ decay amplitude in
  the continuum limit}},
  \href{http://dx.doi.org/10.1103/PhysRevD.91.074502}{\emph{Phys. Rev. D} {\bf
  91} (2015) 074502}, [\href{http://arxiv.org/abs/1502.00263}{{\tt
  1502.00263}}].

\end{thebibliography}\endgroup

\end{document}